\documentclass[floats,aps,showpacs,twocolumn]{revtex4}

\usepackage[dvips]{graphicx}
\usepackage{amsfonts}
\usepackage{amssymb}
\usepackage{float}

\newcommand{\leqa}{\stackrel{<}{\scriptstyle \sim}}
\newcommand{\geqa}{\stackrel{>}{\scriptstyle \sim}}
\newcommand{\ef}{\epsilon_{\text {\tiny F}}}
\newcommand{\ecb}{\overline{\cal E}}
\newcommand{\ect}{\widetilde{\cal E}}
\newcommand{\rmb}{\rho_{\text {\tiny MB}}}

\newcommand{\rhob}{\overline{\rho}}
\newcommand{\rhot}{\widetilde{\rho}}
\newcommand{\sca}{{\cal S}}

\newcommand{\scat}{\widetilde{{\cal S}}}

\newlength{\figwidth}
\begin{document}

\title{Level density of a Fermi gas: average growth and fluctuations}

\author{Patricio Leboeuf and J\'er\^ome Roccia}

\affiliation{Laboratoire de Physique Th\'eorique et Mod\`eles
Statistiques$^*$, B\^at. 100, \\ Universit\'e de Paris-Sud, 91405 Orsay Cedex,
France\\}

\date{\today}

\begin{abstract}
We compute the level density of a two--component Fermi gas as a function
of the number of particles, angular momentum and excitation energy. The
result includes smooth low--energy corrections to the leading Bethe
term (connected to a generalization of the partition problem and
Hardy--Ramanujan formula) plus oscillatory corrections that describe shell
effects. When applied to nuclear level densities, the theory provides a
unified formulation valid from low--lying states up to levels entering the
continuum. The comparison with experimental data from neutron resonances gives
excellent results.
\end{abstract}


\maketitle

Many physical properties of an interacting Fermi gas depend on the number of
available states at a given energy, like for instance the optical and
electrical response of solids, or the reaction rates in nuclear processes. The
most common framework to compute the many--body (MB) density of states (DOS)
is a mean--field approximation, where each (quasi-)particle moves
independently in an average self--consistent potential. In this case, the
energy of the Fermi gas is expressed as the sum of the occupied
single--particle (SP) energies. The computation of the MB DOS is thus reduced
to a combinatorial problem: to count the different ways into which the energy
can be distributed among the particles. A first answer to this question was
given by Bethe \cite{bethe}, who showed that at high excitation energies $Q$
(compared to the SP spacing at Fermi energy $\ef$) and for two types of
fermions (protons and neutrons), the MB density grows like $\rmb (Q) \sim
\exp\left(2 \sqrt{a Q}\right)/Q^{5/4}$. The Fermi gas parameter $a=\pi^2 \rhob
(\ef)/6$ depends only on the average SP DOS at $\ef$, $\rhob (\ef)$.

In practice the parameter $a$ is often used as a fitting parameter. For a
given excitation energy $Q$ and particle number $A$, $a(Q,A)$ is extracted
from the available experimental data. In this way, important deviations from
the independent particle model predictions are observed. Though there are
certainly effects that are beyond that model, our purpose here is to show that
a detailed treatment is able to describe features of individual systems with
good accuracy, therefore providing a solid theoretical basis for
extrapolations to unknown regions and for improvements. Generalizing the
results obtained in Ref.\cite{lmr} to a two--component system of given angular
momentum, we show that Bethe's result can be viewed as the first (smooth) term
of an expansion. The corrections to that term do not enter as corrections of
the $a$--parameter (as was often assumed in the past), but simply as
additional terms in the exponential. A first series of terms are smooth in $A$
and $Q$, and provide higher--order (in inverse powers of the excitation
energy) corrections. Keeping only the first correction generates a uniform
expression, which cancel the divergence produce by the $Q^{-5/4}$ term at low
energies, and therefore make unnecessary the use of composite models (\`a la
Gilbert--Cameron). On top of the smooth contributions are oscillatory terms,
that describe density fluctuations as $A$ varies. These are shell effects,
which turn out to be related to the fluctuations of the total energy of the
system. A detailed description of these fluctuations and of the relevant
energy scales is provided. Finally, a comparison of the results with the
nuclear level density at neutron threshold is made. With a few adjustable
parameters, a very good overall agreement is obtained, with a relative error
$\leqa$ 10\% for the logarithm of the density of the 295 nuclei analyzed.

From a theoretical point of view, Ref.\cite{lmr} and the present work may be
viewed as a generalization to MB systems of the theory developed to describe
the SP DOS \cite{bh,bb,gutz}. Compared to direct generalizations of the SP
Weyl's law and of semiclassical trace formulas \cite{trace}, the statistical
approach employed here has the advantage of leading to simple and physically
meaningful results.

The DOS at energy $E$ of a system composed of $Z$ protons, $N$ neutrons and
with projection $M$ of the angular momentum on some given axis is defined as
\begin{eqnarray} \label{rhodef}
\rmb (E,N,Z,M) = \sum_{\nu} \delta (E-E^{\nu}) \delta (N-A^{\nu}_N) \nonumber \\ 
\delta (Z-A^{\nu}_Z) \ \delta (M - M^{\nu}) \ .
\end{eqnarray}
The index $\nu$ denotes all the possible neutron and proton SP configurations
(of arbitrary number of particles), $A^{\nu}_\lambda = \sum_i
n^{\nu}_{\lambda,i}$ are the neutron and proton number of particles,
respectively ($n^{\nu}_{\lambda,i}=0, 1$ are the corresponding occupation
numbers of the $i$-th SP state, and $\lambda=N$ or $P$); $E^{\nu}=\sum_\lambda
\sum_i n^{\nu}_{\lambda,i} \epsilon_{\lambda,i}$ and $M^{\nu}=\sum_\lambda
\sum_i n^{\nu}_{\lambda,i} m_{\lambda,i}$ are the energy and angular momentum
projection, where $\epsilon_{\lambda,i}$ and $m_{\lambda,i}$ denote the SP
energies and angular momentum projections, respectively.

The conservation of the angular momentum projection is introduced in order to
deal with only a subset of states, those with given total angular momentum
$J$. A standard treatment \cite{bloch,ericson} of this degree of freedom leads
to a DOS of angular momentum $J$ of the form
\begin{equation} \label{rhoj}
\rmb (Q,N,Z,J) = \frac{2J+1}{2\sqrt{2\pi}\sigma^3} {\rm
e}^{-\frac{(J+1/2)^2}{2\sigma^2}} \ \rmb (Q,N,Z),
\end{equation}
where $\rmb (Q,N,Z)$ is the total MB DOS, $Q=E-E_0$ is the excitation energy
measured with respect to the ground state of the system, and $\sigma$ is
the spin cutoff parameter.

For an arbitrary SP spectrum the computation of the density of excited states
is a difficult combinatorial problem for which no exact solution exists. There
is, however, a particular case that can be worked out explicitly: when the SP
spectrum consists of equidistant levels separated by $\delta$ (we assume, for
simplicity, that the neutron and proton spacings are equal). The MB excitation
energies are then given by the sum of two integers corresponding to the total
energy of each of the components, $Q=(j+k) \ \delta=K \ \delta$. Each MB state
characterized by an integer $K$ has a nontrivial degeneracy. The computation
of the degeneracy reduces to the computation of the number of ways into which
the total energy may be distributed among the two components, and of the
different ways the partial energy of each component can be distributed among
its elements. This leads to compute the value of the function $p_2 (K) =
\sum_{j=1}^K p(j) \ p(K-j)$, where $p(j)$ is the partition of $j$ (the number
of ways into which the integer $j$ can be decomposed as a sum of integers). We
are assuming here, to avoid finite size effects, that the excitation energy is
small compared to the Fermi energy of each component. Based on the work of
Hardy and Ramanujan, an exact expression (written as a convergent series) for
$p(j)$ was obtained by Rademacher \cite{hrr}. We have adapted their method
(i.e., the circle method, cf Ref.\cite{andrews}) to obtain an exact formula
for $p_2 (K)$. Putting back the appropriate units, the MB density can be
expressed in terms of $p_2$ as $\rmb /\rhob = 2^{1/4} p_2 (K=\rhob \ Q /2)$,
where $\rhob=\rhob_P + \rhob_N = 2/\delta$ is the total (proton + neutrons) SP
average density. Then, expressing the exact result as an expansion in terms of
$\rhob \ Q$ valid in the range $\rhob^{-1} \ll Q < N \delta, Z \delta$, we
obtain
\begin{equation} \label{rhoeq1}
\rmb(Q,N,Z)/\rhob = \frac{6^{1/4}}{12 \ (\rhob \ Q)^{5/4}} \ {\rm e}^{\sca}
\end{equation}
where the ``entropy'' $\sca=\sca_{eq}$ of the equidistant spectrum is given by
\begin{equation} \label{seq}
\sca_{eq} = 2 \sqrt{\frac{\pi^2}{6} \rhob Q} - \left( \frac{\pi}{36} +
\frac{15}{16 \pi} \right) \frac{\sqrt{6}}{\sqrt{\rhob Q}} + \left(
\frac{35}{96} + \frac{\pi^2}{432} \right) \frac{1}{\rhob Q}
\end{equation}
plus ${\cal O}((\rhob \ Q)^{-3/2})$ corrections that can be computed but are
not given here. The prefactor in Eq.(\ref{rhoeq1}) and the first term of the
expansion (\ref{seq}) reproduce Bethe's formula \cite{bethe}. The additional
terms provide further smooth corrections of higher order in inverse powers of
the excitation energy. Though Eq.(\ref{seq}) represents an asymptotic
expansion, we find that an accurate uniform approximation is obtained by
keeping only up to the $1/\sqrt{\rhob Q}$ term that kills, when $Q \rightarrow
0$, the divergence produced by the $(\rhob Q)^{-5/4}$ in the prefactor. It is
also interesting to note that the correct coefficients of the correction terms
in (\ref{seq}) are obtained through the expansion of the exact result, whereas
a saddle point approximation of the sum involved in $p_2 (K)$ leads to wrong
coefficients.

The previous expression describes in detail the MB DOS for a SP spectrum made
of equidistant levels. However, it is clearly unphysical in most situations. A
generic SP spectrum contains {\sl fluctuations}, which are manifested at the
scale of the average distance between levels, but also on much larger scales
(see Ref.\cite{seville} for a review). What is missing in Eq.(\ref{rhoeq1})
are the fluctuations in the MB density induced by the SP fluctuations. In this
respect, one may consider Eq.(\ref{rhoeq1}) as the MB analog of the Weyl or
Wigner-Kirkwood expansions.

It remains to compute the MB level density for an arbitrary SP spectrum,
including fluctuations. The way to do it was shown, for a single--component
gas, in Ref.\cite{lmr}. The method uses a saddle point approximation of the
inverse Laplace transform of the MB density. We have adapted that calculation,
following similar lines, to treat Eq.(\ref{rhodef}), that includes two
components and angular momentum conservation. The result may be written under
the form of Eqs.(\ref{rhoj}) and (\ref{rhoeq1}), but with the entropy in the
latter equation given by
\begin{equation} \label{sfin}
\sca = \sca_{eq} + \frac{1}{T} \left[ \ect (N,Z,0) - \ect (N,Z,T) \right] \ .
\end{equation}
The parameter $T$ is the temperature, connected to the excitation energy $Q$
through the usual relation $Q = \pi^2 \rhob T^2/6= a \ T^2$. $\ect (N,Z,T) =
\sum_\lambda \int d \epsilon \ \rhot_\lambda (\epsilon) \ \epsilon \
f(\epsilon,\mu,T)$ is the fluctuating part of the energy of the system at
temperature $T$ and chemical potential $\mu \sim \ef$ fixed, neglecting
temperature variations, by the particle--number conditions $N \sim Z =
\int^{\ef} d \epsilon \ \rho_{\lambda} (\epsilon)$. The function $\rho_
\lambda (\epsilon) = \sum_j \delta (\epsilon - \epsilon_{\lambda,j})$ is the
SP density of the component $\lambda$, and $\rhot_\lambda (\epsilon) =
\rho_\lambda (\epsilon) - \rhob_\lambda (\epsilon)$ its fluctuating part.
$\ect (N,Z,0)$ is thus the fluctuating part of the ground state energy of the
system. Finally, $\sca_{eq}$ in Eq.(\ref{sfin}) is given by Eq.(\ref{seq}),
with $\rhob$ the total average SP density of the system at Fermi energy. In
fact, for an arbitrary spectrum the saddle point technique does not allow to
derive the terms of order $(\rhob Q)^{-1/2}$ and higher in $\sca_{eq}$. The
corrections obtained from an equidistant spectrum are thus conjectured to
provide a good approximation to the corrections of the smooth part of an
arbitrary system, but the validity of this statement has to be confirmed. An
explicit numerical verification of its validity for a two--dimensional
one--component system was done in Ref.\cite{lmr}.

The function $\ect (N,Z,T)$ presents oscillations when $N$ or $Z$ are varied,
in contrast to the more gentle variations as a function of $T$ (a detailed
description of the fluctuations and of their $T$--dependence is given below).
The MB level density contains now two types of terms: some that vary smoothly,
and others that fluctuate as the number of particles varies. The result
presented above contains the dominant smooth and oscillatory terms. In the
derivation of Eqs.(\ref{rhoj}), (\ref{rhoeq1}) and (\ref{sfin}) we have
neglected other terms (for instance, the chemical potentials and $T$ have
small corrections that depend on $J$, and thus strictly speaking the
factorization (\ref{rhoj}) of the angular momentum is not exact, etc). A
detailed account of the derivation will be given elsewhere.

It is remarkable that the MB level density at excitation energy $Q$ depends
explicitly on the ground--state energy fluctuations $\ect (N,Z,0)$. A
convenient way to analyze the behavior of the fluctuating part of the entropy
$\scat = \left[ \ect (N,Z,0) - \ect (N,Z,T) \right]/T$ is through a
semiclassical theory. The result is an expression for $\scat$ written as a sum
over all the classical periodic orbits of the mean field potential. The main
conclusions that can be drawn from that expression are now listed. To be
specific, we consider the particular case of an atomic nucleus of $Z$ protons
and $N$ neutrons: (a) as the mass number $A=Z+N$ varies at a fixed excitation
energy $Q$, $\scat$ presents oscillations of characteristic period $\delta A
\approx (\pi/3) A^{2/3}$, that are independent of $Q$; (b) when $Q$ varies at
fixed particle number $A$, $\scat$ does not present similar oscillations, but
rather gentle variations; (c) the typical amplitude $\sigma_{\scat}$ of
$\scat$ at given $(Q,A)$ depends on the dynamical properties of the classical
dynamics (integrable or chaotic); (d) since the mean field dynamics of most
nuclei is well approximated by a regular motion \cite{seville}, then the
behavior of the typical amplitude of $\scat$ is given, to a first
approximation, by those of a regular dynamics, that we now detail; (e) using
the definition of the temperature $T=\sqrt{Q/a}$ with $a=\pi^2 \rhob/6 \approx
A/15$ MeV$^{-1}$ \cite{bm}, we find that there is only one relevant
temperature scale in the variation of $\sigma_{\scat}$ with $T$, given by $T_c
\approx 4/A^{1/3}$ MeV; for convenience we also introduce $T_{\delta} = (2
\pi^2 \rhob)^{-1}\approx 1.3/A$ MeV (the temperature associated to the SP mean
level spacing), and $g=T_c/T_{\delta} \approx 3 A^{2/3}$; (f) at low
temperatures, $\sigma_{\scat} \approx \sqrt{T/T_{\delta}}$; the typical
amplitude of shell effects in the MB DOS therefore increases from 0 at $T=0$
up to $\sim \sqrt{g}$ at $T \sim T_c$; (g) at temperatures of order $T_c$ the
amplitude is maximal, and starts to decrease for $T > T_c$; (h) in the limit
$T \gg T_c$ the typical amplitude tends to zero as $\sigma_{\scat} \approx
\sqrt{g/6} \ T_c/T$; using the previous values of $T_c$ and $g$ this gives
$\sigma_{\scat} \approx 2 \sqrt{2} /T$ MeV; we thus predict a slow power--law
decay of the amplitude of shell effects at high temperatures.

We now turn to a direct application of the previous results to experimental
data. Though it will be important to make a systematic analysis of the
validity of Eqs.(\ref{rhoj}), (\ref{rhoeq1}) and Eq.(\ref{sfin}) and of their
predictions at different energies and mass numbers, we restrict here to a
comparison with slow neutron resonances, which have been experimentally
studied for a large number of nuclei \cite{RIPL2}. The excitation energies of
neutron resonances coincide with the neutron binding energies, $Q = Q_n
= S_n (N,Z)$, whose values are in the range $6-8$ MeV for most nuclei.
This corresponds to a temperature $T_n \approx 8/\sqrt{A}$ MeV. According to
the previous results, the typical amplitude of the fluctuations depends on
temperature, with a maximum at $T \approx T_c$. At neutron resonances the
ratio $T_n /T_c \approx 2/A^{1/6}$. From $A=30$ to $A=250$, this ratio varies
from $1.13$ to $0.8$. We thus find that at excitation energies $Q \approx
Q_n$, the temperature is very close to $T_c$; shell effects are maximal. We
expect a typical value of $\scat (Q,N,Z)$ very close to its maximum $\sim
\sqrt{g} \approx \sqrt{3} A^{1/3}$ (this varies from 5.4 to 11 in the previous
range of $A$). In contrast, in the same particle--number range the first
correcting term (proportional to $(\rhob \ Q)^{-1/2}$) in the smooth expansion
(\ref{seq}) varies from 0.32 to 0.11. That term, and the following ones in the
expansion, can thus be neglected at $Q\approx Q_n$.

To make a comparison with experiments we need the different quantities
involved in the theoretical expressions. One possibility is to compute them
from a particular model. In our case, however, in order to avoid
model--dependent features and to make a direct test of our predictions we
prefer to extract as much information as possible from experimental data. For
each nucleus, the excitation energy at neutron threshold $Q = Q_n = S_n (N,Z)$
is taken from the experimental value of $S_n (N,Z)$, and $T_n (N,Z) =
\sqrt{Q_n/a}$. For $\ect (N,Z,T_n)$ we have no experimental data available. We
compute it as follows. Semiclassically, $\ect (N,Z,T_n)$ is written as a sum
over the periodic orbits $p$ (and repetitions) of the mean field potential
\cite{seville}. The analysis of the temperature dependence of that sum and of
the main contributing orbits leads to the approximation $\ect (N,Z,T_n)
\approx \overline{\kappa}_n \ \ect (N,Z,0)$, where $\overline{\kappa}_n$ is
the average over the shortest periodic orbits $p$ of the function $\kappa
(x_p) = x_p/\sinh (x_p)$, where $x_p = 3 \pi \ell_p A^{1/3} T_n/(4 \ef)$ and
$\ell_p$ is the length of the periodic orbit $p$ measured in units of the
nuclear radius (notice the mass number and temperature dependence of $x_p$).
For each nucleus, $A$ and $T_n$ are given and the average $\overline{\kappa}_n
= \overline{\kappa} (N,Z,T_n)$ is computed. In practice, the average is
estimated using the shortest periodic orbits of a spherical cavity of radius
$R = 1.2 A^{1/3}$fm.

The expression of the entropy $\sca$ in Eq.(\ref{rhoeq1}) takes now the form
\begin{equation} \label{sexp}
\sca (Q_n,N,Z) = 2 \sqrt{a \ Q_n} + (1 - \overline{\kappa}) \ \ect (N,Z,0) /T_n \ .
\end{equation}
Finally, $\ect (N,Z,0)$ and $a$ are required. $\ect (N,Z,0)$ can be obtained
from the experimental value of the ground--state energy. It corresponds to the
fluctuating part of the nuclear binding energy, that we compute by subtracting
from the 1995 Audi--Wapstra compilation \cite{aw} the liquid drop expression
$\ecb = a_v A - a_s A^{2/3} - a_c Z^2/A^{1/3} - a_A (N-Z)^2/A$, using the
parameters (from Ref.\cite{prsz}) $a_v=15.67, \ a_s=17.23, \ a_A=23.29$, and
$a_c=0.714$ (all in MeV; we have moreover excluded the pairing term). This
parametrization produces a fluctuating part whose average (over $A$) $\langle
\ect (N,Z,0) \rangle_A \approx 0$. However, the determination of the average
of the fluctuating part is a delicate question that deserves a careful
discussion. Due to the discrete variation of the chemical potential as the
mass number varies, one can verify that generically $\langle \ect (N,Z,0)
\rangle_A$ is non-zero. We have therefore added to the fluctuating part a term
$b A + c$, where $b$ and $c$ are two constants. Equation (\ref{sexp}) thus
depends on three constants, $a$, $b$, and $c$, that we fix by minimizing the
root mean square error with respect to the experimental value of the density,
$\sca_{exp}$ (obtained by computing $\sca$ from Eqs.(\ref{rhoj}) and
(\ref{rhoeq1}) when $\rmb (Q,N,Z,J)$ is the experimental DOS, $J$ the ground
state angular momentum and $\sigma^2 \approx 0.15 a A^{2/3} T_n$
\cite{RIPL2}). The result is $a=A/10.42$ MeV$^{-1}$, $b=-0.019$ MeV and
$c=7.9$ MeV. The comparison is made in Fig.1. The experimental values
$\sca_{exp}$ shown on the top part are to be compared with the ``theoretical''
entropies plotted in the middle part. A clear overall agreement is observed.
For most nuclei, the relative error in the lower panel is smaller than 10\%,
with some remaining structure as a function of $A$, and larger deviations for
closed shells (we suspect that this is due to our very schematic estimate of
$\overline{\kappa}_n$).

The precision of the present calculation, with only three adjusted parameters,
is comparable to the best fits obtained nowadays. We can in fact make the
comparison more precise by noticing that Eq.(\ref{sexp}) can be approximated,
using an effective value of $a$, by $\sca (Q_n,N,Z) \approx 2 \sqrt{a_{ef} \
Q_n}$, where $a_{ef}=a \ [1+ \ect (N,Z,0) (1-\overline{\kappa})/Q_n]$. Under
this form, Eq.(\ref{sexp}) is quite similar to one of the best
phenomenological formulas studied so far, proposed by Ignatyuk and
collaborators \cite{RIPL2,ist}.

To conclude, we have derived an explicit formula for the MB DOS of a two
component Fermi gas of fixed angular momentum. The results were applied to the
particular case of nuclear level densities, where precise predictions for the
smooth dependence and shell fluctuations as a function of excitation energy
and mass number were made. Good agreement between theory and experiment in the
region of neutron resonances is found. Although it was derived within an
independent particle model, the comparison with experiments shows that the
final result is probably of more general validity and includes, through the
energy fluctuations, effects like pairing. Going to high excitation energies,
the main prediction is the decay of shell effects when $Q \geqa Q_n$ (with a
power--law tail). However, that prediction is valid for closed systems. Before
proceeding in that direction, the theory should be improved to include finite
size effects (e.g. a finite number of nucleons) as well as the influence of
the continuum.

\noindent $^*$ Unit\'e de recherche associ\'ee au CNRS. We acknowledge fruitful
discussions with C. Schmit. This work was supported by grants ACI Nanoscience
201, ANR NT05-2-42103, ANR-05-Nano-008-02 and the IFRAF Institute.

\begin{figure} [H]
\includegraphics[width=8.5cm,height=7.4cm]{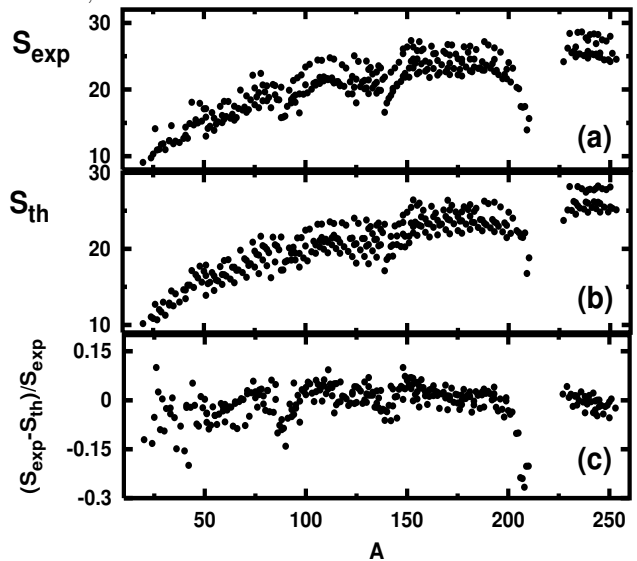} 
\caption{Entropy $\sca$ as a function of the mass number $A$ for nuclear level
densities at neutron threshold. (a) experimental values; (b) theoretical
prediction; (c) relative error.}
\end{figure}

\end{document}